\documentclass[prd,superscriptaddress,nofootinbib]{revtex4-2}

\usepackage{color}
\usepackage{bm}
\usepackage[utf8]{inputenc}
\usepackage{makecell}
\usepackage{subfig}
\usepackage{graphicx}  
\usepackage{dcolumn}   
\usepackage{bm}        
\usepackage[english]{babel}
\usepackage{booktabs, multirow, array}
\usepackage[bottom]{footmisc}
\usepackage[table]{xcolor}
\usepackage{float}
\usepackage{makecell}
\usepackage{dblfnote}
\usepackage{mathtools}
\usepackage{amsmath}
\DFNalwaysdouble 
\usepackage[colorlinks]{hyperref}
\definecolor{LinkColor}{rgb}{0.75, 0, 0}
\definecolor{CiteColor}{rgb}{0, 0.5, 0.5}
\definecolor{UrlColor}{rgb}{0, 0, 0.75}
\hypersetup{linkcolor=LinkColor}
\hypersetup{citecolor=CiteColor}
\hypersetup{urlcolor=UrlColor}

\usepackage[normalem]{ulem} 

\begin{document}
 \title{\bf Regularized Stable Kerr Black Hole: Cosmic Censorships, Shadow and Quasi-Normal Modes}
\author{Rajes Ghosh}
\email{rajes.ghosh@iitgn.ac.in}
\affiliation{Indian Institute of Technology, Gandhinagar-382355, Gujarat, India}
\author{Mostafizur Rahman}
\email{mostafizur.r@iitgn.ac.in}
\affiliation{Indian Institute of Technology, Gandhinagar-382355, Gujarat, India}
\author{Akash K Mishra}
\email{akash.mishra@icts.res.in}
\affiliation{International Centre for Theoretical Sciences, Tata Institute of Fundamental Research, Bangalore 560089, India}


\begin{abstract}

Black hole solutions in general relativity come with pathologies such as singularity and mass inflation instability, which are believed to be cured by a yet-to-be-found quantum theory of gravity. Without such consistent description, one may model theory-agnostic phenomenological black holes that bypass the aforesaid issues. These so-called regular black holes are extensively studied in the literature using parameterized modifications over the black hole solutions of general relativity. However, since there exist several ways to model such black holes, it is important to study the consistency and viability of these solutions from both theoretical and observational perspectives. In this work, we consider a recently proposed model of regularized stable rotating black holes having two extra parameters in addition to the mass and spin of a Kerr solution. We start by computing their quasi-normal modes under scalar perturbation and investigate the impact of those additional parameters on black hole stability. In the second part, we study shadows of the central compact objects in $M87^*$ and $Sgr\, A^*$ modelled by these regularized black holes and obtain stringent bounds on the parameter space requiring consistency with Event Horizon Telescope observations.

\end{abstract}

\maketitle
\section{Introduction}

Einstein's theory of general relativity (GR) predicts the existence of compact dark objects such as black holes (BHs). Under the assumption of strong energy condition, collapsing matter in GR leads to spacetime singularities \cite{hawking_ellis_1973, Wald:1984rg, PhysRevLett.14.57}. These set of results are known as Hawking-Penrose singularity theorems that suggest a generic existence of singularities in BH solutions of GR. At the location of a spacetime singularity, various curvature scalars diverge, indicating breakdown of the underlying gravity theory and its predictive power. It is widely believed that at small enough length scales, general relativity would be replaced by a quantum theory of gravity, which will ultimately resolve this issue. However, in the absence of a fully consistent quantum gravity framework, one may consider a phenomenological approach in which the known BH solutions are modified to regularize the central singularity with a non-singular core. In general, such approaches are highly non-unique and theory agnostic in nature. Several interesting regularized BH solutions (both static and rotating) have been proposed over years \cite{Dymnikova:1992ux, Modesto:2008im, Bonanno:2000ep, Hayward:2005gi, Platania:2019kyx, Bambi:2013ufa, Azreg-Ainou:2014pra, PhysRevD.89.104017, Eichhorn:2021etc, Delaporte:2022acp}. Such solutions contain one or more additional parameters over the usual hairs, namely the mass and spin of a Kerr BH. Usually these regularized BHs possess Cauchy horizons with nonzero surface gravity causing an exponential growth of perturbations at its vicinity known as the mass inflation singularity \cite{PhysRevD.101.084047, Carballo-Rubio:2019nel, PhysRevD.41.1796, PhysRevLett.67.789, Bhattacharjee:2016zof}. However, the recently proposed regularized BH solutions in Ref.\cite{Carballo-Rubio:2022kad} (spherically symmetric) and in Ref.\cite{Franzin:2022wai} (rotating) circumvent both the issues of singular core and mass inflation instability. Due to its astrophysical relevance, we focus on this regularized rotating BH geometry which we refer as regularized stable Kerr BH and study various  properties in the subsequent sections.\\

\noindent
As discussed previously, the approach of regularizing the singularity is highly non-unique and as a result, it leads to a plethora of regularized BHs of different types. In the absence of a unique approach, an effective way to check the viability of these models could be to perform various consistency tests. In this paper, we explore the regularized BH solution presented in Ref.\cite{Franzin:2022wai} and attempt to constrain the deviations from Kerr BH by theoretical and observational studies. The central theme of this article is of two folds. Firstly, inspired by seminal works of Regge, Wheerle and Vishweswara, we check the stability of such solutions by analysing the scalar quasi-normal mode (QNM) spectrum \cite{PhysRev.108.1063, PhysRevLett.24.737, Vishveshwara:1970zz, PhysRevLett.29.1114, Leaver:1985ax, PhysRevD.34.384}. Ensuring the stability of a BH solution is essential not only from a mathematical perspective but also due to its astrophysical relevance. For a comprehensive review on the subject of BH QNMs and stability, we refer the reader to Refs. \cite{Berti:2009kk, Kokkotas:1999bd, Konoplya:2011qq}.
If the regularized BH under consideration is found to be unstable in certain region of the parameter space, one can rule out those regions. This is a very powerful method to check the viability of any classical BH solution and constrain any additional parameter present in the theory. Conversely, any well behaved BH solution must also be stable against small perturbation in the entire range of the parameter space. One of the aims of this work is to look for any instabilities present in this BH spacetime. Our study suggests that the regularized BH under consideration is a stable configuration against scalar perturbation. As a result, to constrain possible deviation from Kerr solution, we resort to observational techniques, which corresponds to the second part of our work. 
\\ 

\noindent
The bending of light by a strong gravitational field is one of the fascinating features of the spacetime curvature, which gives rise to the existence of spherical/circular null orbits around BHs. As far as observational implications are concerned, these null orbits further lead to BH shadow, a dark patch around BHs seen by a distant observer. The recent observations of $M87^*$ and $Sgr\, A^*$ shadows by the Event Horizon Telescope (EHT) collaboration~\cite{EventHorizonTelescope:2019dse, EventHorizonTelescope:2019ths, EventHorizonTelescope:2019pgp, EventHorizonTelescope:2019ggy, EventHorizonTelescope:2022xnr, EventHorizonTelescope:2022xqj} are the first direct detection of isolated BHs, providing us with an unprecedented platform to test the viability of modified Kerr solutions. However, since the EHT observations are in strong agreement with the Kerr paradigm, any deviation from Kerr present in the theory are highly constrained~\cite{Jusufi:2022loj, Hendi:2022qgi, Banerjee:2022iok, KumarWalia:2022aop, Banerjee:2022bxg, Banerjee:2022jog, Shaikh:2022ivr, Li:2022eue}. Assuming the central compact objects in $M87^*$ and $Sgr\, A^*$ are modelled by the regularized stable Kerr BHs, we compare the angular shadow sizes with the EHT observations and obtain upper bounds on the additional parameters.\\

\noindent
The rest of the article is organized as follows: In Sec.\ref{Sec_2} we briefly review the regularized stable Kerr solution and discuss some of its properties. In Sec.\ref{Sec_3} we discuss the consistency conditions on the mass profile suggested in Ref. \cite{Franzin:2022wai}. This in some sense represents a study similar to the weak cosmic censorship~\cite{WALD1974548, Hubeny:1998ga, Sorce:2017dst, Ghosh:2021cub}. The computation of scalar quasi-normal modes and stability analysis is presented in Sec.\ref{Sec_4}. In Sec.\ref{Sec_5}, we study the shadow cast by this BH and constrain various model parameters requiring consistency with EHT observations. Finally, we conclude with a brief discussion of main results and possible future extensions of our work.\\

\noindent
Notations and conventions: In this work we follow the mostly positive signature convention $(-, +, +, +)$ for the metric. Indices referring the four dimensional spacetime are represented by Greek letters. Unless specified otherwise, we also set the fundamental constants to unity, i.e., $c = 1 = G$.

\section{Constructing the metric} \label{Sec_2}
An astrophysical BH with Arnowitt-Deser-Misner (ADM) mass $M$ and spin $a<M$ is well described by the Kerr metric, which is an exact solution of Einstein field equations. However, it is infected with three major issues: (i) the central singularity where various curvature scalars blow up and GR looses its predictive power, (ii) the instability originating due to the mass inflation at the Cauchy horizon, and (iii) the existence of closed timelike curves in the spacetime. Over years, several attempts have been made to get rid of one such issue at a time by introducing some modified phenomenological metrics. However, in a recent work \cite{Franzin:2022wai}, the authors used the following ``inner-degenerate'' regularized stable Kerr metric that is free from all these issues,

\begin{align} \label{g}
ds^2 = \mathcal{C}(r,\theta) \Bigg[-\left(1-\frac{2\, m(r)\, r}{\Sigma(r,\theta)}\right) \mathrm{d} t^{2}-\frac{4\, a\, m(r)\, r\, \sin ^{2} \theta}{\Sigma(r,\theta)}\, \mathrm{d} t\, \mathrm{d} \phi\, +\, &\frac{\Sigma(r,\theta)}{\Delta(r)}\, \mathrm{d} r^{2}\,  + \\ \nonumber
&\Sigma(r,\theta)\, \mathrm{d} \theta^{2}+\frac{A(r,\theta) \sin ^{2} \theta}{\Sigma(r,\theta)}\, \mathrm{d} \phi^{2}\Bigg] .
\end{align}

\noindent
This metric can be easily obtained from from the metric presented in Eq.(16) of Ref.\cite{Azreg-Ainou:2014pra}. And, the special case where $\mathcal{C}(r,\theta) \equiv 1$ corresponds to the well-known G$\ddot{\text{u}}$rses-G$\ddot{\text{u}}$rsey metric \cite{Gurses:1975vu}. Here, we have used the notations:  $\Sigma(r,\theta)=r^{2}+a^{2} \cos ^{2} \theta,\, \Delta(r)=r^{2}-2\, m(r)\, r+a^{2},\, A(r,\theta)=\left(r^{2}+a^{2}\right)^{2}-\Delta\,  a^{2} \sin ^{2} \theta$, and $m(r)$ is the mass profile introduced to remove the mass inflation instability at the Cauchy horizon $r_{-}$. Note that the locations of the horizons $r_{+}$ and $r_{-} < r_{+}$ are given by the real positive roots of $\Delta(r) = 0$. On the other hand, the conformal factor $\mathcal{C}(r,\theta)$ is there to tame the central curvature singularity of the Kerr metric.\\

\noindent
We are interested in investigating the QNM stability and the shadow observables of this modified Kerr BH and put constraints on various parameters of the metric. To keep our analysis as general as possible, we shall work with the mass profile $m(r)$ and the conformal factor $\mathcal{C}(r,\theta)$ unspecified, and only fix them when necessary. However, in the same spirit of Ref.\cite{Franzin:2022wai}, we impose two minimal criteria on the conformal factor in order to avoid any additional singularity: $\mathcal{C}(r,\theta) > 0$ everywhere, and to keep the spacetime indistinguishable from Kerr metric to a distant observer: $\mathcal{C}(r,\theta) \to 1 + \mathcal{O}(r^{-n})$ with $n \geq 2$. The second condition implies that the ADM mass and the specific angular momentum of the black hole are $M = \text{lim}_{r \to \infty} m(r)$ and $a$, respectively. \\ 

\noindent
Also, we follow the same efficient way of Ref.\cite{Franzin:2022wai} for parameterizing $m(r) = F(r; r_{\pm})$ in terms of the event horizon at $r_{+} = M+\sqrt{M^2-a^2}$ and a degenerate Cauchy horizon at $r_{-}=a^2\, \left[ M+(1-e)\sqrt{M^2-a^2}\right]^{-1}$:

\begin{equation}
    m(r) = M\, \frac{r^2+\alpha\, r + \beta}{r^2+\gamma\, r+\mu}\, .
\end{equation}

\noindent
To remove the mass inflation singularity, we should choose $\beta \neq 0$ and $\alpha \neq \gamma$ so that the inner horizon is degenerate and surface gravity is zero there. Following Ref.\cite{Franzin:2022wai}, the coefficients are given by

\begin{align}
\alpha & =\frac{a^4+r_{-}^3 \, r_{+}-3\, a^2\,  r_{-}\left(r_{-}+r_{+}\right)}{2\,  a^2\, M}\, , \\
\beta & =\frac{a^2\left(2\, M-3\,  r_{-}-r_{+}\right)+r_{-}^2\left(r_{-}+3\, r_{+}\right)}{2\, M}\, , \\
\gamma & =2\, M-3\, r_{-}-r_{+}\, , \\
\mu & =\frac{r_{-}^3\, r_{+}}{a^2}\, .
\end{align}

\noindent
Then, for a given value of $e$, the extremal limit $r_{-} \to r_{+}$ remains same as the Kerr case (namely $a=M$), which we shall refer as ``a-extremality". Moreover, to ensure $0< r_{-} < r_{+}$, we need to choose the parameter $e$ such that $0 \neq e < 2$. The metric in Eq.(\ref{g}) reduces to a conformal Kerr BH by setting $e=0$. On the other hand, $e \to 2$ is the ``e-extremal" limit for any fixed $a \leq M$. \\

\noindent
One may ask whether the above parameterization of the mass profile $m(r)$ is consistent and physical. It may seem hard to answer this question in the absence of the hitherto unknown field equation that may support the phenomenological metric in Eq.(\ref{g}) as a solution. However, there are two ways to tackle this issue. Assuming the metric to be a non-vacuum solution of GR, one may write down an effective energy-momentum tensor and check whether various energy conditions are satisfied. It is discussed in detail in Ref.\cite{Franzin:2022wai}. The authors conclude that for $\alpha < \gamma$, not only weak but null and dominant energy conditions are met. Though the strong energy condition is always violated, one should not be too alarmed with it since the metric may not be a solution to Einstein's field equations. For the same reason and to keep our analysis theory-agnostic, we will not put any prior constrain on the parameter space of $e$ except what is dictated by the regularity of the metric itself, namely  $e \in ( -3 - 3/ \sqrt{1-a_*^2},\, 2]$ with $a_* := a/M$ \cite{Franzin:2022wai}. Another way to check the consistency of the mass profile is to consider its evolution under test particles absorption, which we shall discuss in the next section.

\section{Consistency criterion on the mass profile} \label{Sec_3}
As discussed in the previous section, we want to impose a specific parameterization on the mass profile $m(r) = F(r; r_{\pm})$. This is only possible upto the a-extremal ($a=M$) configuration of the BH given by Eq.(\ref{g}). Now consider the process in which such a BH with $ a \leq M$ absorbs a test particle with energy $E << M$ and angular momentum $L << M^2$. This may change both the ADM mass and the angular momentum of the initial black hole in such a way that the final configuration has $(a\, M + L) > (M+E)^2$ and $r_{\pm}$ becomes complex. As a result, occurrence of such a process puts the viability of the aforesaid parameterization in question. Thus, we want to put some constraints on $m(r)$ that can prevent such difficulties.\\

\noindent
The above inequality holds if the particle's energy satisfy the upper bound $E < E_{max} = aL/(M^2+a^2)$. On the other hand, such a particle will be absorbed by the BH if it does not come across any turning points before crossing the event horizon at $r_{+}$. Neglecting back reactions and using geodesic equations, the entering condition can be translated to a lower bound on the test particle's energy: $E \geq E_{min} = \left(a\, \sigma\, L \right)/\left[2 M r_+ + a^2(\sigma-1)\, sin^2\theta \right]$, where $\sigma = m(r_+)/M$. Thus, to prevent such a process we must require, $E_{min} \geq E_{max}$. It implies the consistency condition: $\sigma \geq r_+^2/M^2$, that is strictest for all allowed values of $\theta \in [0,\pi)$. Therefore, if the mass profile satisfy the inequality $m(r_+) \geq r_+^2/M$, the BH will not absorb those particle that could have made the parametrization inconsistent in the final configuration.\\

\noindent
For the mass profile given in Ref.\cite{Franzin:2022wai}, this consistency criteria is not satisfied if the initial BH is sub-extremal with $a<M$. Therefore, one may choose a different mass profile which satisfies the consistency criteria under test particle approximation. However, in the same spirit of \cite{Sorce:2017dst}, it is reasonable to expect that the situation improves and the apparent inconsistency in the parametrization of Ref.\cite{Franzin:2022wai} goes away as one properly takes into account the back reaction effects. In contrast, if we start with an a-extremal BH, the consistency condition becomes $m(r_+) \geq M$, which is satisfied by the choice of $m(r)$ given in Ref.\cite{Franzin:2022wai} even under test particle approximation.\\

\noindent
Though it looks similar to the calculation done to check the validity of the weak cosmic censorship (WCC) conjecture for BHs~\cite{WALD1974548, Hubeny:1998ga, Sorce:2017dst, Ghosh:2021cub}, there is a subtle difference. WCC protects the predictive power the underlying gravity theory, it is necessary to censor out those BH solutions in which there is a possibility of destroying the horizon and exposing the central singularity by test particle absorption. However, for the metric in Eq.(\ref{g}), there is no such curvature singularity inside the event horizon. Thus, the inequality derived above is not a strict necessity, it is just to make the aforesaid parameterization consistent. If we use different parameterization, we would have got a different criteria on $m(r)$. However, for the purpose of this paper, we shall stick to the same parameterization for the mass profile as given in Ref.\cite{Franzin:2022wai}.

\section{Scalar QNM spectrum} \label{Sec_4}
Note that, so far we have not specified any particular form for the conformal factor $\mathcal{C}(r,\theta)$ or the mass profile $m(r)$ except putting some conditions on them. However, to calculate the QNM spectrum and shadow of this BH in later sections, we have to specify their functional forms. For the mass profile, we consider the same form as given in Eqs.(28-32) of Ref.\cite{Franzin:2022wai}.\\

\noindent
However, we choose a different conformal factor $\mathcal{C}(r,\theta) = 1+b/r^{2z}$ (with $b>0$, so that the conformal factor is non-zero everywhere) than what are proposed in Ref.\cite{Franzin:2022wai}. This choice for $\mathcal{C}(r,\theta)$ is motivated by two facts. Firstly, it is much simpler to work with and performs the job that it is designed for, namely it regularizes the central curvature singularity of the Kerr metric if $z \geq 2$. In addition, such a choice of the conformal factor makes the scalar wave equation separable, which we shall exploit in finding the QNM spectrum. However, for the purpose of this paper, we shall consider the simplest case $z=2$.\\

\begin{figure}
	\centering
	\minipage{0.48\textwidth}
	\includegraphics[width=\linewidth]{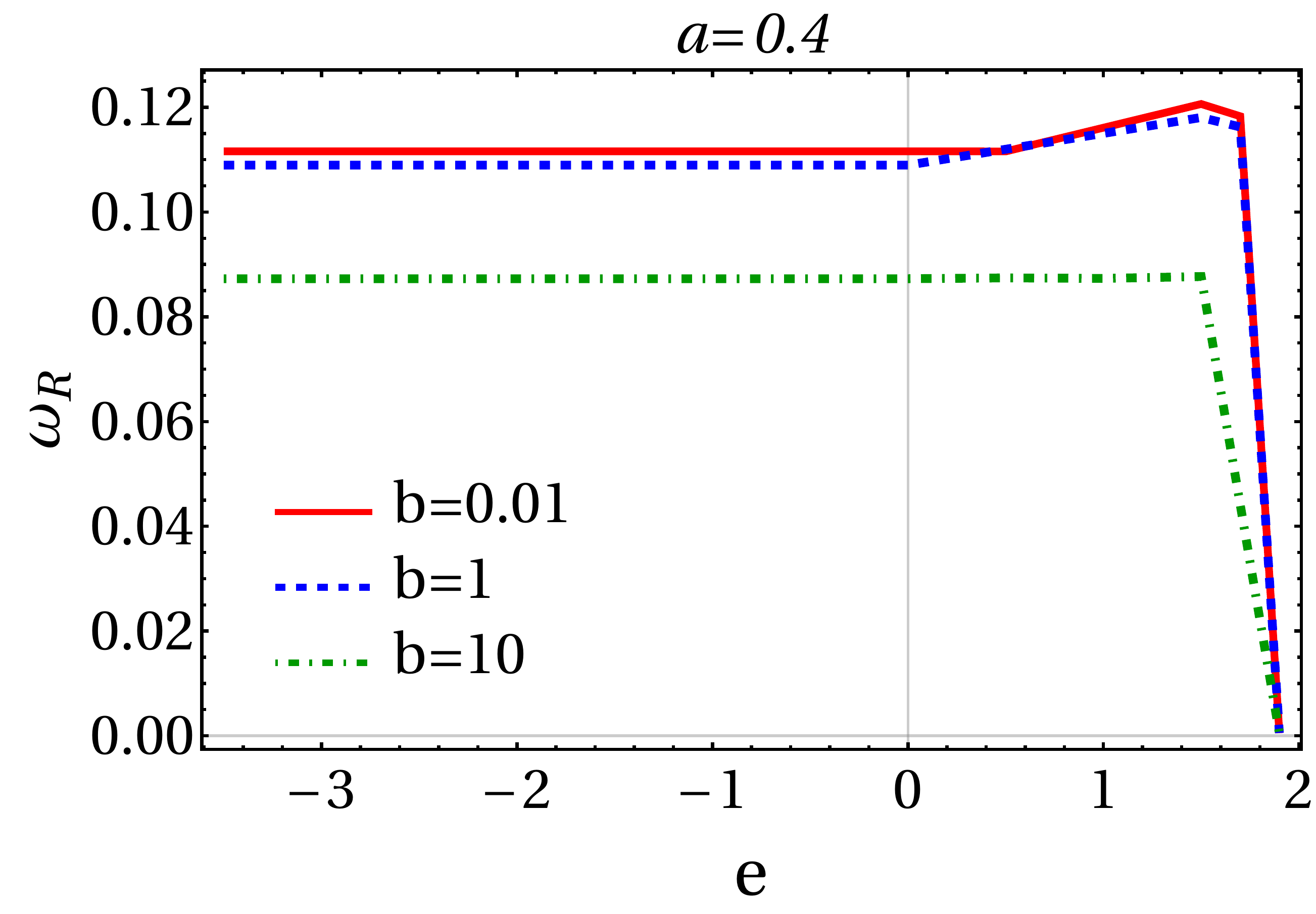}
	\endminipage\hfill
	\minipage{0.48\textwidth}
	\includegraphics[width=\linewidth]{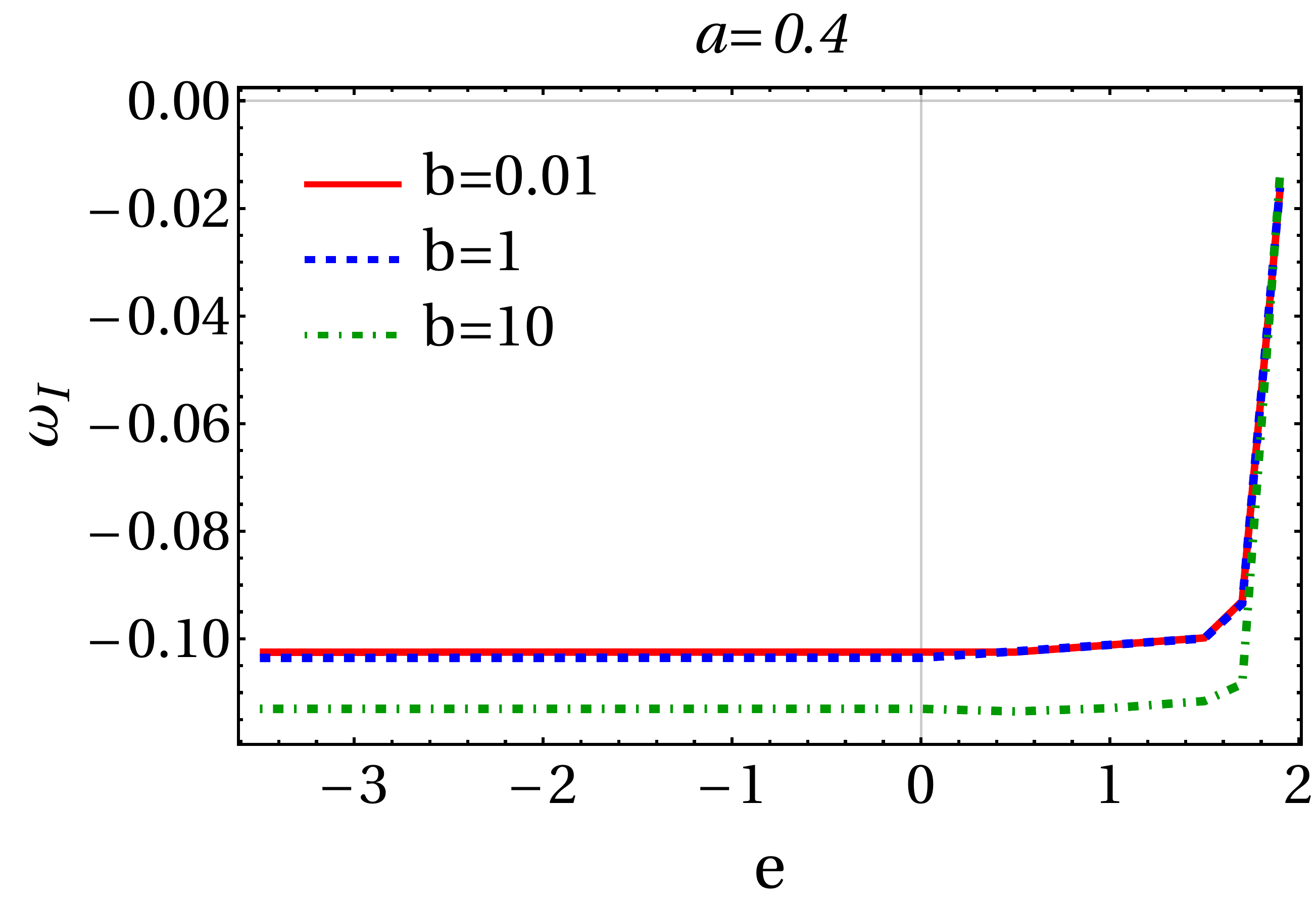}
	\endminipage
	\hfill
	\minipage{0.48\textwidth}
	\includegraphics[width=\linewidth]{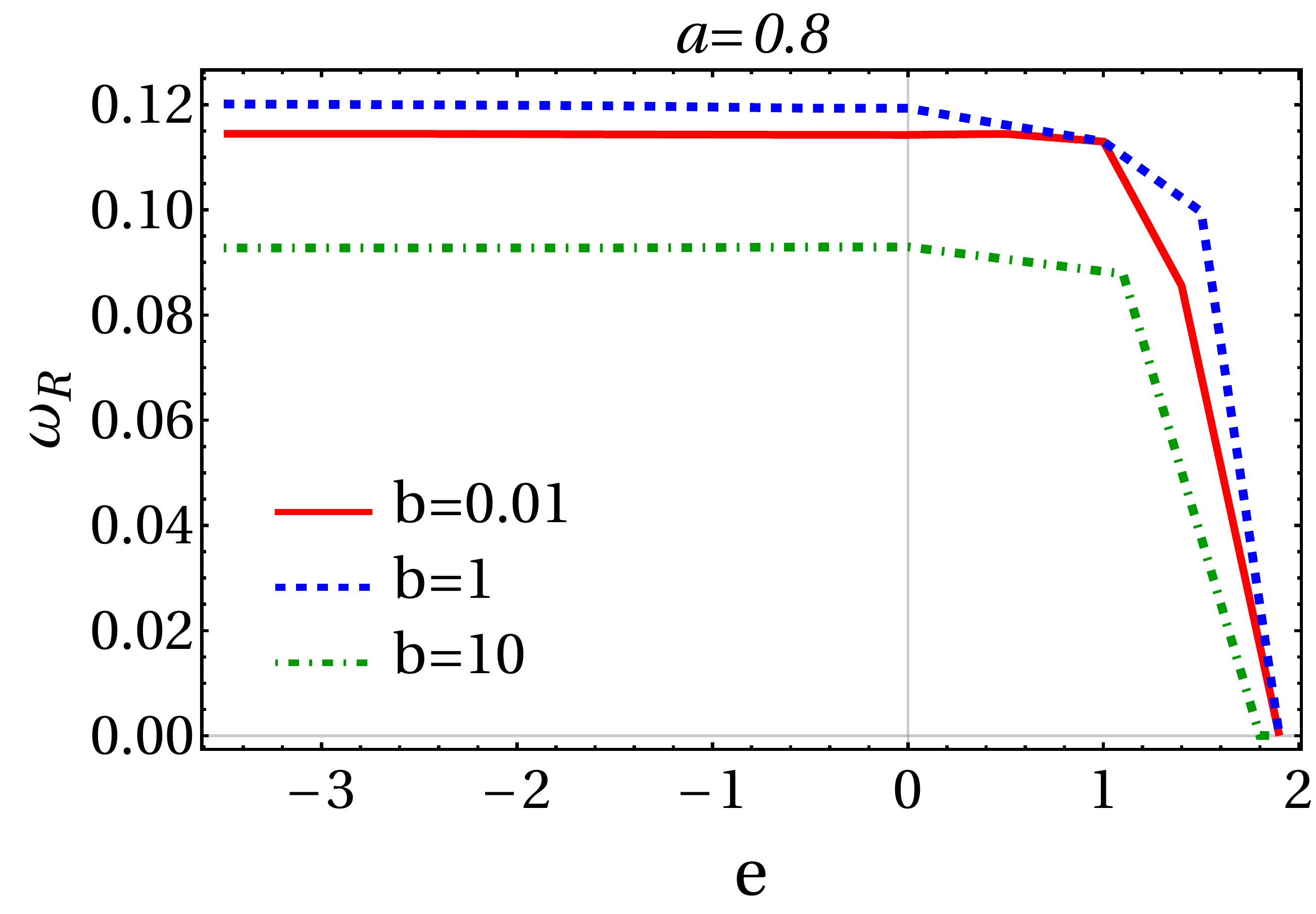}
	\endminipage\hfill
	\minipage{0.48\textwidth}
	\includegraphics[width=\linewidth]{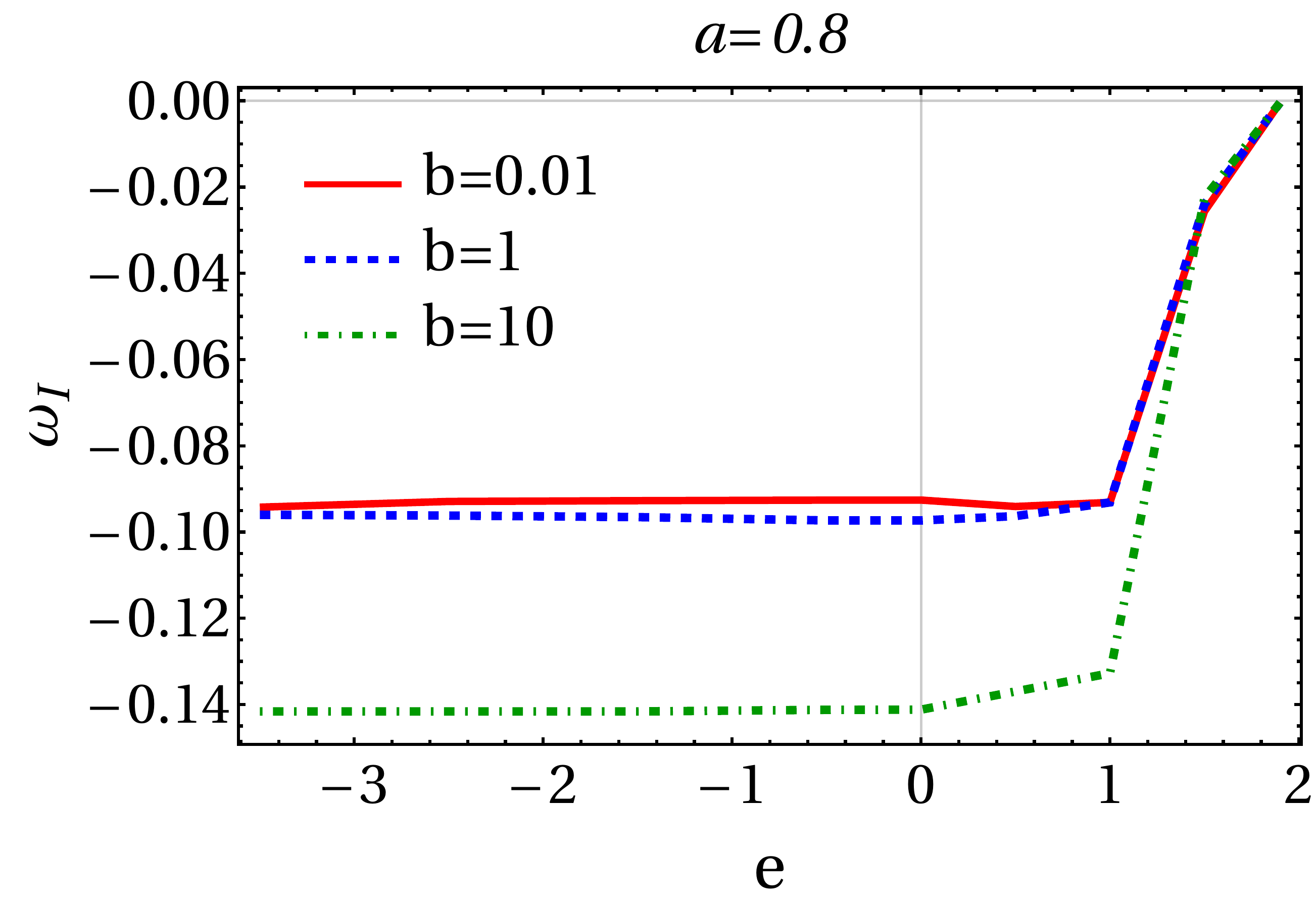}
	\endminipage
	\caption{The variation of both real (left panel) and imaginary parts (right panel) of $\ell = 0$ scalar QNM modes are shown. The parameters are taken to be: $M=1$, $a \in \{0.4,0.8\}$, $e \in ( -3 - 3/ \sqrt{1-a^2},\, 2]$, and $b \in \{0.01,1,10\}$.}\label{fig_stability_Scalar}
\end{figure}

\noindent
 Consider that a massless scalar field $\Phi(x^\mu)$ perturbs the the stationary and axisymmetric BH geometry given in \ref{g} with $\mathcal{C}(r,\theta) =\psi(r)= 1+b/r^{4}$, and $m(r)$ as prescribed in Ref.\cite{Franzin:2022wai}. Owing to the symmetries of the background spacetime, we can decompose the field as $\Phi(x^\mu)=  R_{lm}(r)\, S_{lm}(\theta) e^{-i\,( \omega\, t-\, m\, \phi)}$, where $S_{lm}(\theta)$ is the spheroidal harmonics which satisfies the angular Teukolsky equation \cite{1973ApJ...185..635T},

\begin{equation}\label{ang_Teuk}
\left[\frac{d}{dx}\left(1-x^2\right)\frac{d}{dx}+\left(E_{lm}+a^2\, \omega^2\, x^2-\frac{m^2}{1-x^2}\right)\right]S_{lm}(x)=0,
\end{equation} 

\noindent
with $x=\cos\theta$ and $E_{lm}$ is the separation constant. Note that even in the presence of conformal factor $\psi(r)$, we can cast the radial equation into the standard Teukolsky form \cite{1973ApJ...185..635T}:

\begin{equation} \label{rad_teuk}
	\left[\frac{d}{dr}\, \Big(\Delta _1\, \frac{d}{dr} \Big)+	 \left(\frac{  K^2}{\Delta _1}-\lambda_{lm}\, \psi \right)\right] R_{lm}(r)=0 ,
\end{equation}

\noindent
by introducing a function $\Delta_1 (r) =\psi(r)\,  \Delta(r)$. Here, $\lambda_{lm}=E_{lm}+a^2\, \omega ^2-2\, a\, m\, \omega $ and $K=\psi(r)\left[\omega  \left(a^2+r^2\right)-a\, m\right]$. The QNMs are the solution of the above equations with ingoing boundary conditions at the event horizon and outgoing boundary conditions at infinity, 
\begin{align}
R_{lm}(r\rightarrow r_+)\sim e^{-i\omega r_*}\qquad \text{and} \qquad R_{lm}(r\rightarrow \infty)\sim e^{i\omega r_*}\ ,
\end{align}

\noindent
where $r_*$ is the tortoise coordinate defined by $dr/dr_* := \Delta_1(r)/(r^2+a^2)$. The radial function $R_{lm}$, subject to the above QNM boundary conditions, can be expressed as follows \cite{Rahman:2019uwf, Rahman:2020guv},
\begin{equation}\label{frobenious}
    R_{lm}=\frac{u_{lm}(z)}{r}\, e^{i\, B_1(r)}\, z^{-2\, i\,  B_2(r_+)}\ .
\end{equation}

\begin{figure}[ht]
	\centering
	\minipage{0.48\textwidth}
	\includegraphics[width=\linewidth]{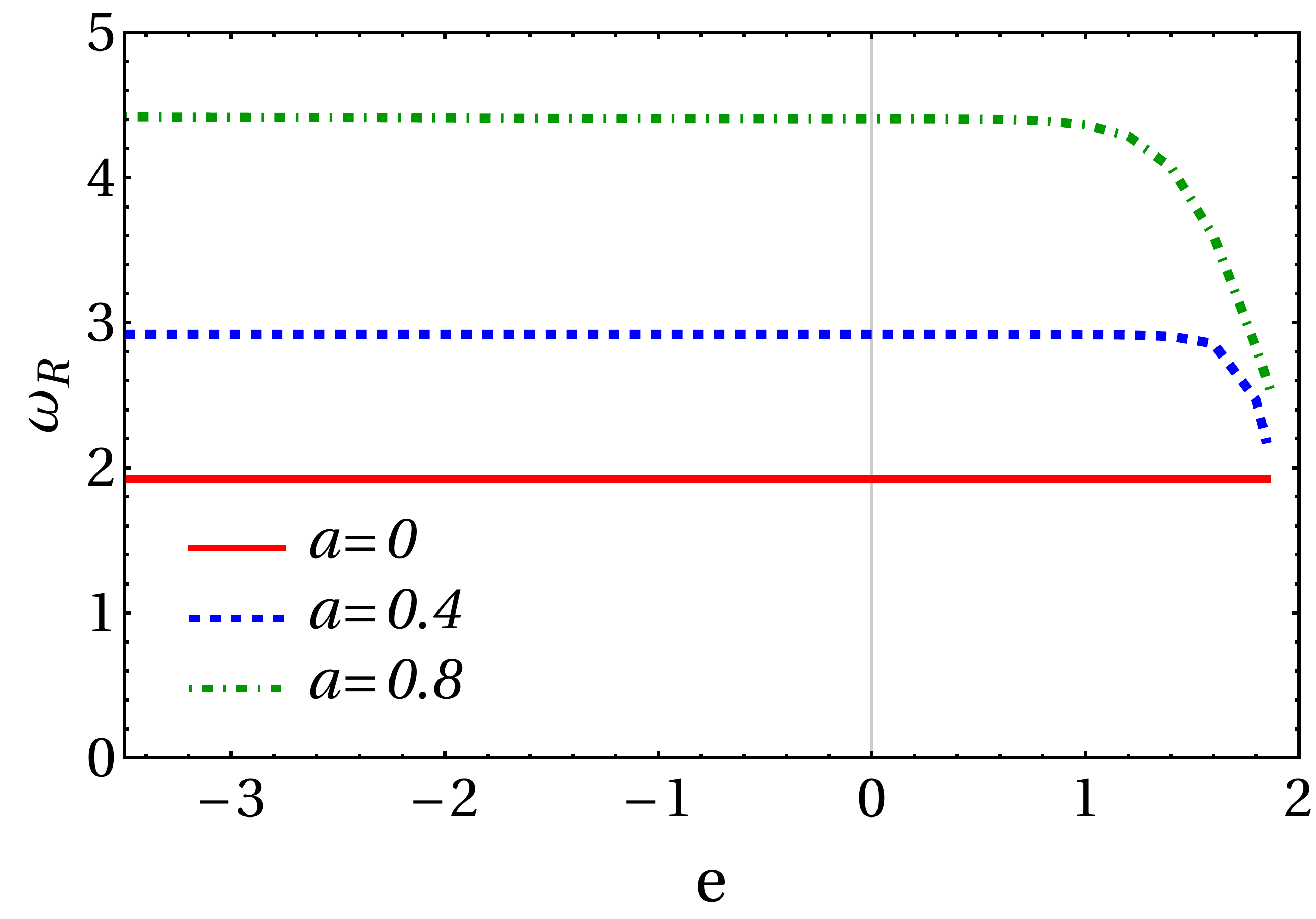}
	\endminipage\hfill
	\minipage{0.48\textwidth}
	\includegraphics[width=\linewidth]{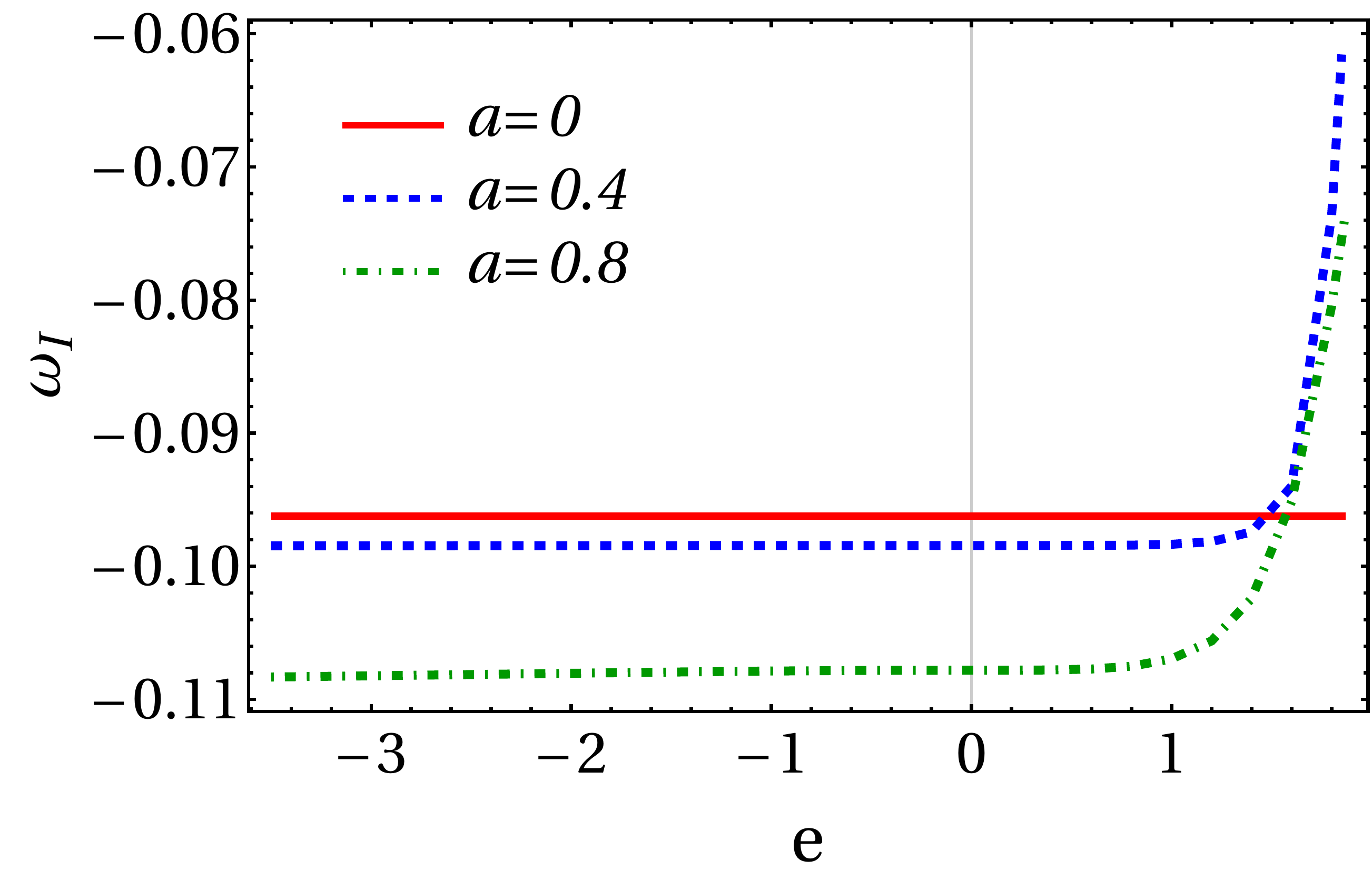}
	\endminipage
	\caption{In this figure, we represent the real and imaginary part of the QNM computed analytically in the eikonal limit with $\ell=10$. The parameters are taken to be: $M=1$, $a \in \{0,0.4,0.8\}$, $e \in ( -3 - 3/ \sqrt{1-a_*^2},\, 2]$, and $b=0.1$.}\label{fig_Lyapunov}
\end{figure}

\noindent
Here, $z=(r-r_+)/(r-r_-)$, $B_1(r)=K(r)/\Delta_1(r)$, and  $B_2(r_+)=K(r_+)/\Delta'_1(r_+)$. Replacing the above expression in Eq.(\ref{rad_teuk}), we obtain an equation for $u_{lm}(z)$. Note that both the radial and angular equation depends on $\omega$ and $\lambda_{lm}$. Thus, to obtain the desired QNMs, we need to solve these two equations simultaneously. We use the \texttt{Black Hole Perturbation Toolkit} \cite{BHPToolkit} to solve the angular Teukolsky equation in Eq.(\ref{ang_Teuk}), and  \texttt{QNMSpectral} package \cite{Jansen:2017oag} to solve the equation for $u_{lm}(z)$ numerically. For given values of $(a, b, e, \ell)$, the output of this analysis is the set of QNM frequencies and separation constants. The real and imaginary part of the scalar QNMs are shown in Fig.[\ref{fig_stability_Scalar}] for various choice of parameters $(e, b, a)$ for the case of $\ell = 0$ fundamental mode. As it can be seen from Fig.[\ref{fig_stability_Scalar}], the imaginary part of QNM remains negative for the entire range of the additional parameters $e$ and $b$, indicating the regularized black hole to be stable in the entire parameter space.
Furthermore, we interestingly find that with increasing values of the conformal parameter ($b$), the imaginary part of the scalar QNM frequency decreases for the entire range of $e$. As the conformal factor is introduced to tame the central curvature singularity, we can conclude that BHs with regular cores are more stable as compared to the singular ones. Also,  Fig.[\ref{fig_stability_Scalar}] clearly demonstrates the dependence of QNMs on the parameter $e$. For a given value of $b$, the decay time of the perturbation increases (or, stays the same for non-rotating case) with increasing values of $e$.\\
Although we have specified a form of the conformal factor in the previous section, we shall see that for the study of the BH shadow, no particular form for $\mathcal{C}(r,\theta)$ needs to be assumed. It is because for any choice of the conformal factor $\mathcal{C}(r,\theta)$, null geodesic equations are separable and take the following form,

\noindent
Because of the complicated structure of the perturbation equations in a general curved spacetime, it is often difficult to obtain any analytical results. In such a scenario, one resorts to numerical techniques, as we did previously. However, in some limited cases, it is indeed possible to solve the perturbation equation analytically. One such approximation is the eikonal limit or geometric optics limit. We have cross-verified our results for eikonal modes (presented in Fig.[\ref{fig_Lyapunov}])  with analytical Lyapunov methods \cite{2010cfsm.bookC, Cardoso:2008bp, nagle2004fundamentals}, which are in excellent agreement with each other. In the eikonal limit ($\ell >> 1$ keeping $ m = \ell)$, the QNMs can be interpreted in terms of various properties of the equatorial null geodesics trapped at the photon sphere at $r = r_c$. The real parts of the QNMs are proportional to the angular frequency $\Omega_c$ of these null geodesics at the photon sphere. In contrast, the imaginary part is related to the Lyapunov exponent $\Lambda_{\textrm{Lya}}$, which dictates the instability timescale of the equatorial null geodesics at the photon sphere. Thus, the eikonal QNMs take the form \cite{Cardoso:2008bp}: 

\begin{equation}\label{lya}
    \omega_{n\ell}=\ell\, \Omega_c-i\,  (n+\frac{1}{2})\, |\Lambda_{\textrm{Lya}}|\, ,
\end{equation}

\noindent
where $n$ is the overtone number. The expression for the $\Omega_c$ and $\Lambda_{\textrm{Lya}}$ can be written in terms of the metric coefficients as \cite{Rahman:2018oso},
\begin{equation}
    \begin{aligned}
    & \Omega_c =\frac{g_{t\phi}+g_{tt}\, y}{g_{\phi\phi}+g_{t\phi}\, y}\Bigg\rvert_{r = r_c}\, ,\\
    & \Lambda_{\textrm{Lya}}^2 =\frac{g_{t\phi}^2-g_{tt}\, g_{\phi\phi}}{g_{rr}}\left\{\frac{g_{tt}\, g''_{\phi\phi}-g''_{tt}\, g_{\phi\phi}-2\, y \left(g''_{tt}\, g_{t\phi}-g_{tt}\, g''_{t\phi}\right)}{2\, g_{tt}\left(g_{\phi\phi}+y\,  g_{t\phi}\right)^2}\right\}\Bigg\rvert_{r = r_c}\, ,
    \end{aligned}
\end{equation}
where $g_{\mu\nu}$ denotes the metric coefficient given in Eq.(\ref{g}), prime ($'$) denotes derivative with respect to $r$ and $y$ is the impact factor, the expression of which can be written as follows \cite{Rahman:2018oso},

\begin{equation}
    y=-\frac{g_{t\phi}}{g_{tt}}+\sqrt{\left(\frac{g_{t\phi}}{g_{tt}}\right)^2-\frac{g_{\phi\phi}}{g_{tt}}}\Bigg\rvert_{r = r_c}\, .
\end{equation}

\noindent
To summarize, this section considers the scalar perturbation of the regularized stable Kerr metric and computed the QNM spectrum both analytically and numerically. We found that both for small and large values of $\ell$, the imaginary part of the QNM modes are always negative, implying the stability of the underlying BH in the entire parameter space, namely $b>0$ and $e \in ( -3 - 3/ \sqrt{1-a_*^2},\, 2]$ with $a_* = a/M$. Thus, it is not possible to constrain these extra parameters from the stability analysis. As a result, we resort to shadow observations in order to obtain any possible bounds on the additional parameters, which is discussed in the next section.

\section{Bounds from Shadow Observations} \label{Sec_5}
BH shadow provides us with a powerful tool to constrain various parameters of the metric under consideration. In this section, we shall study the shadow cast by the regularized stable Kerr BH and compare it with the EHT observations for $M87^*$ and $Sgr\, A^*$ shadows \cite{EventHorizonTelescope:2019dse, EventHorizonTelescope:2019pgp, EventHorizonTelescope:2022xnr}. The goal is to restrict the parameter space of these regularized BHs such that their shadow sizes match with EHT observations.\\

\noindent
Although we have specified a form of the conformal factor in the previous section, we shall see that for the study of the BH shadow, no particular form for $\mathcal{C}(r,\theta)$ needs to be assumed. It is because for any choice of the conformal factor $\mathcal{C}(r,\theta)$, null geodesic equations are separable and take the following form,

\begin{align} \label{geod}
\mathcal{C}^2(r,\theta)\, \Sigma^2(r,\theta)\, \dot{r}^2 = -Q\, \Delta(r) + \Big[ (r^2+a^2)\, E - a\, L_z \Big]^2 := R(r), \\ 
\text{and}\quad \mathcal{C}^2(r,\theta)\, \Sigma^2(r,\theta)\, \dot{\theta}^2 = Q - \Big[ \frac{L_z}{\text{sin}\theta} - a\, E\, \text{sin}\theta \Big]^2 := \Theta(\theta)\ .
\end{align}

\noindent
Here, $E$ and $L_z$ are two constants of motion associated to photon's energy, and the $z$-components of the angular momentum. The separability implies the existence of another constant of motion $Q$, namely the Carter constant. Now, to construct the BH shadow, we require the presence of a photon region filled with unstable spherical null geodesics satisfying $R(r_p) = R'(r_p) = 0$ with $R''(r_p) < 0$. Solving these two equations simultaneously, we obtain two useful quantities, 

\begin{equation}
\xi := \frac{L_z}{E} = \frac{1}{a} \Big[ r_p^2 +a^2 -\frac{2\, r_p\, \Delta(r)}{r_p-M(r_p)} \Big],\, \, \text{and} \quad \eta := \frac{Q}{E^2} =  \frac{4\, r_p\, \Delta(r_p)}{\left[ r_p - M(r_p) \right]^2}\, ,
\end{equation}

\noindent
where $M(r) = m(r) + r\, m'(r)$ with the mass profile $m(r)$ given in Ref.\cite{Franzin:2022wai}. However, one can not put any arbitrary values of $r_p$ in the above set of equations. It is because the spherical null geodesics are confined only in a finite portion of the spacetime outside the BH known as the photon region. This region is represented by the inequality $\Theta(\theta, \xi, \eta) \geq 0$. At any fixed value of $\theta$, this inequality is satisfied when $r_p$ varies between two end points $\left[r^p_{-}, r^p_{+}\right]$ where equality holds true. In particular, at the equatorial plane, these end points correspond to light rings. In fact, the existence of at least one light ring outside the ergoregion is assured by a theorem proved in Ref.\cite{Ghosh:2021txu}.\\

\noindent
One can use $(\xi, \eta)$ to span the observer's sky with suitably defined celestial coordinates $(\alpha, \beta)$. For a distant observer, these coordinates are given in terms of the inclination angle $\theta_i$:

\begin{equation} \label{celes}
\alpha = - \xi\, \text{cosec}\theta_{i},\, \, \text{and}\, \, \, \, \beta = \pm \sqrt{\eta - \Big(\xi\, \text{cosec}\theta_{i} - a\, \text{sin}\theta_i\Big)^2}\ .
\end{equation}

\begin{figure}[ht]
\begin{center}
\includegraphics[width=0.4\textwidth]{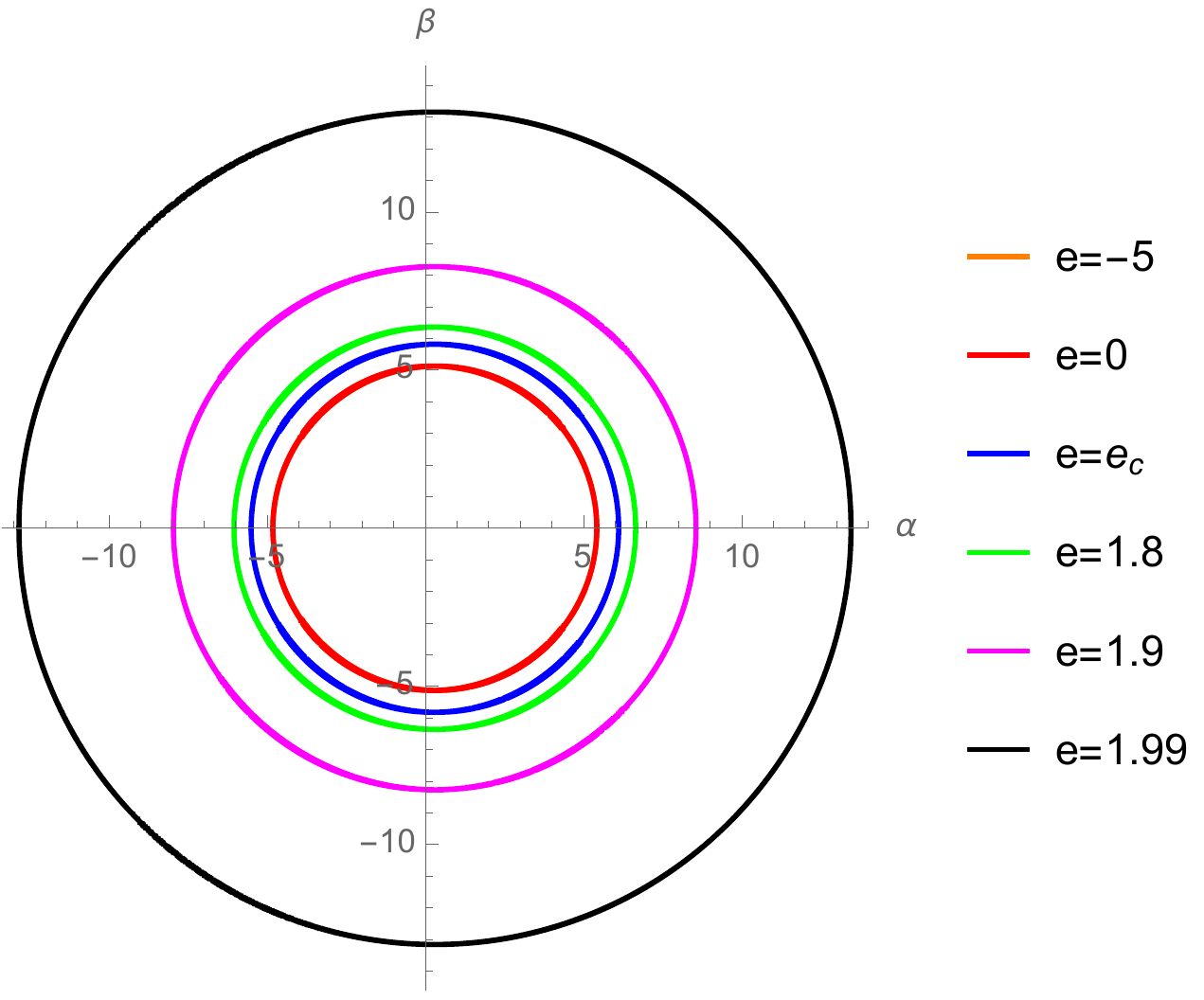}
\caption {This figure shows the shadow corresponding to metric (Eq. \ref{g}) if used to model $M87^*$ BH shadow ($M=1$) for $a_* = 0.5$ and different values of $e$. Beyond a critical value $e > e_c \approx 1.75$, the shadow size becomes bigger than $45\, \mu as$ and becomes inconsistent with EHT observations\cite{EventHorizonTelescope:2019dse}.}
\label{shad}
\end{center} 
\end{figure}

\noindent 
For a Schwarzschild BH, the shadow boundary is a circle of radius $R_{sh} = 3\sqrt{3}\, M$. In contrast, rotation makes the size of a Kerr shadow varies away from $R_{sh}$. Usually, such variation is negligible if the BH is not rotating very rapidly. However, in our case, the shadow of a regularized stable Kerr BH depends on an extra parameter $e$. For a fixed values of $(M, a)$, the BH shadow will have different angular sizes as $e$ varies in the range $( -3 - 3/ \sqrt{1-a_*^2},\, 2]$ (with $a_* = a/M$) as prescribed in Ref.\cite{Franzin:2022wai}. For example, see Fig.[\ref{shad}] in which $M87^*$ shadow sizes (in $M=1$ unit) are shown for $a_* = 0.5$ and different choices of $e$ values.\\

\begin{center}
\begin{table}[ht!]
\begin{tabular}{|c|c|c|c|c|c|c|}
\Xhline{1\arrayrulewidth}
Central BHs  &  Mass  $( M/M_\odot)$  &  Spin ($a_*$)   &  Distance ($D$)                            &  Inclination ($\theta_i$)                        &  Shadow size ($\theta_d$)                        &   References                       
\\\Xhline{1\arrayrulewidth}	
$M87^*$	       &    $6.5 \times 10^9 $   &	 $0.5-0.94$	   &	$16.8\, \text{Mpc}$	&	$ 17 ^{\circ}$               & 	$42 \pm 3\, \mu$as        &   \cite{EventHorizonTelescope:2019dse, EventHorizonTelescope:2019pgp}	
\\\Xhline{1\arrayrulewidth}
$Sgr\, A^*$      &    $4.0 \times 10^6$                       &	 $0.44$	   &	$8\, \text{kpc}$	        &	$< 50 ^{\circ}$             & 	$51.8 \pm 2.3\, \mu$as  &   \cite{EventHorizonTelescope:2022xnr, Kato_2010}	  
\\\Xhline{1\arrayrulewidth}
\end{tabular}
\caption{EHT observational data for $M87^*$ and $Sgr\, A^*$ shadows.}
\label{T1}
\end{table}
\end{center}

\begin{figure}[ht]
  \subfloat[For $-5 \leq e \leq 0$]{
	\begin{minipage}[c][1\width]{
	   0.3\textwidth}
	   \centering
	   \includegraphics[width=1.1\textwidth]{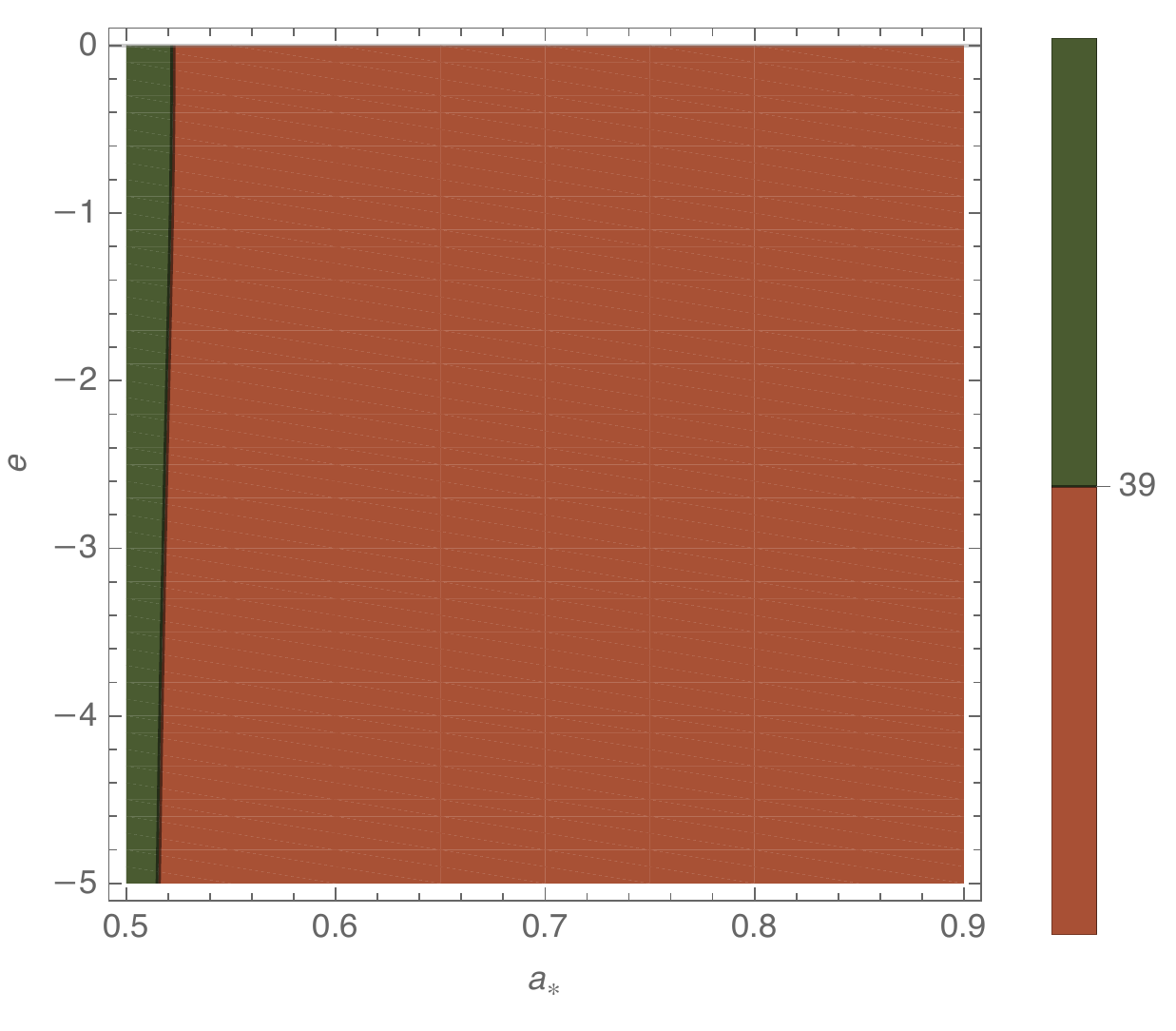}
	\end{minipage}}
 \hfill 	
  \subfloat[For $0 \leq e \leq 1.68$]{
	\begin{minipage}[c][1\width]{
	   0.3\textwidth}
	   \centering
	   \includegraphics[width=1.1\textwidth]{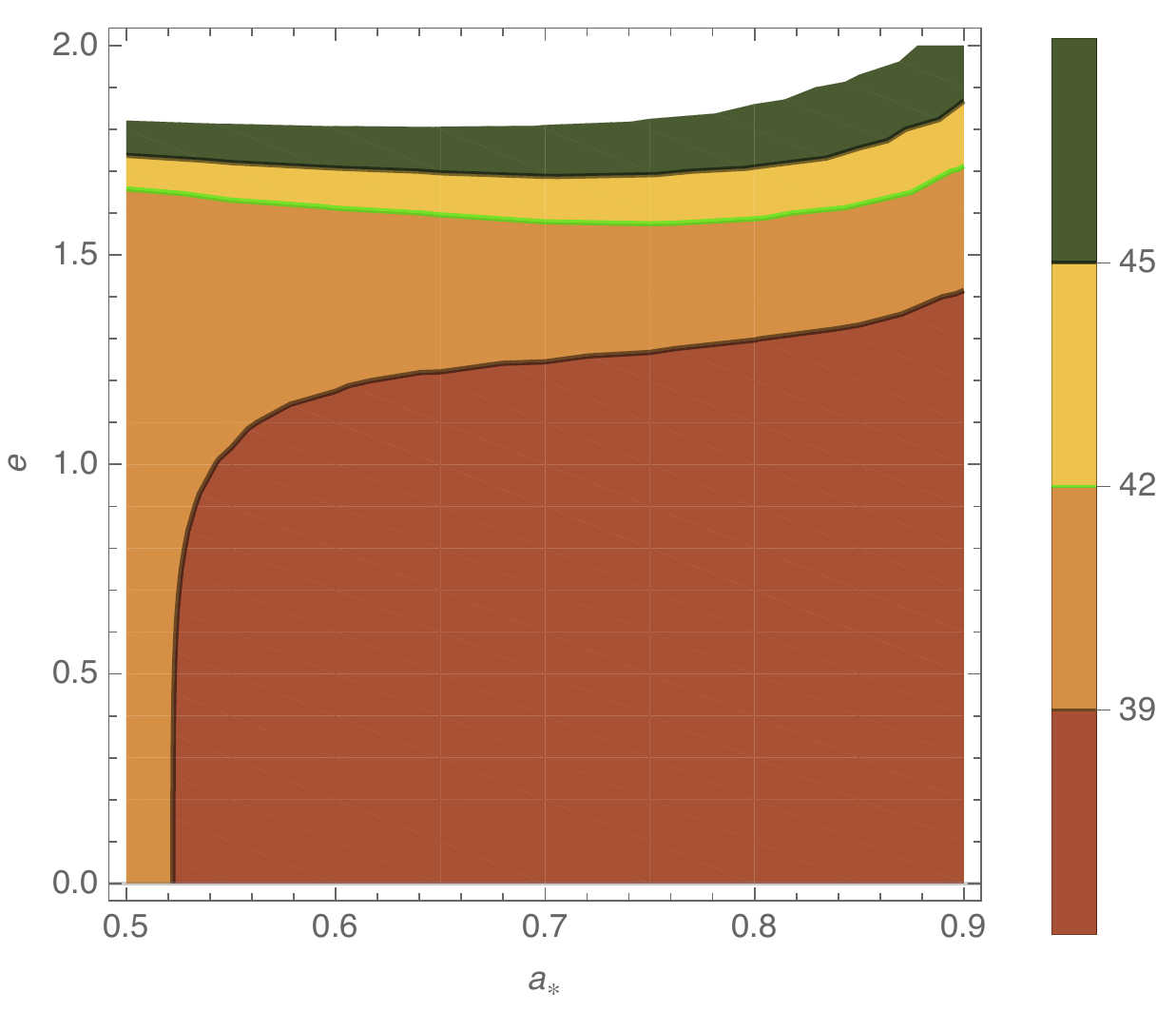}
	\end{minipage}}
 \hfill	
  \subfloat[For $1.68 \leq e \leq 2$]{
	\begin{minipage}[c][1\width]{
	   0.3\textwidth}
	   \centering
	   \includegraphics[width=1.1\textwidth]{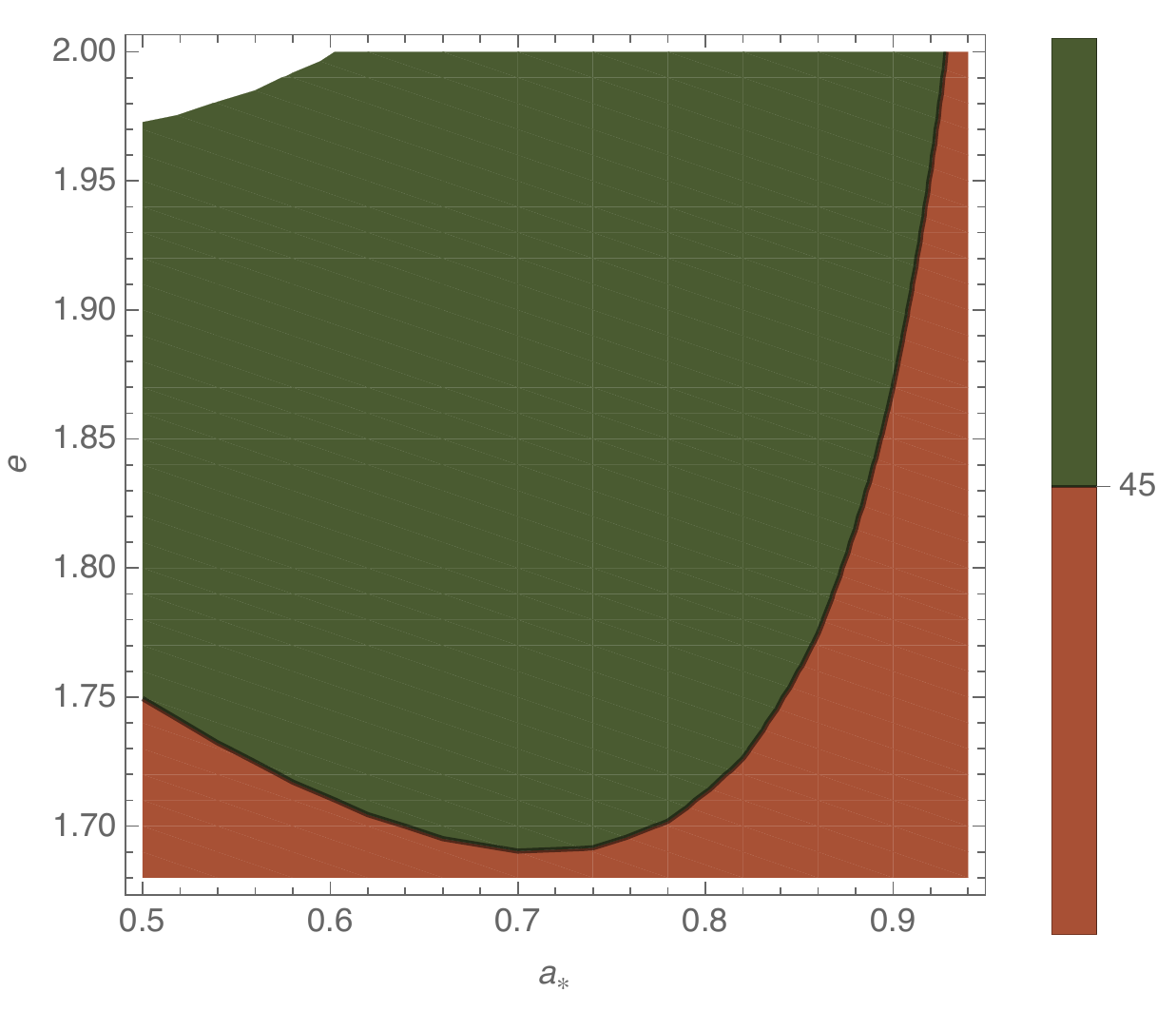}
	\end{minipage}}
\caption{Contour plots for angular size of $M87^*$ shadow if modelled by the metric in Eq.(\ref{g}). Values of $a_* \in [0.5,0.94]$, and $e \in (-5, 2)$ for which the shadow size is less than $39\, \mu$as or greater than $45\, \mu$as are disallowed. Values of $b$ do not effect the shadow size.}
\label{M87plt}
\end{figure}

\noindent
Thus, for fixed values of $(M, a_*, \theta_i, D)$, there can be a range of $e$-values for which the shadow will be inconsistent with the angular size $\theta_d$ \cite{Hioki:2009na, Kumar:2020owy} as observed by EHT, check Table-\ref{T1}. For this purpose, let us first consider the $M87^*$ shadow. In this case, since the angular momentum parameter consistent with observation can take a range of values $a_* \in [0.5,0.94]$, we have shown below the variation in shadow size using contour plots Fig.[\ref{M87plt}]. From these plots, it is very easy to identify the disallowed regions which are inconsistent with EHT observations. \\

\begin{figure}[ht]
  \subfloat[For $-5 \leq e \leq 0$]{
	\begin{minipage}[c][1\width]{
	   0.3\textwidth}
	   \centering
	   \includegraphics[width=1.1\textwidth]{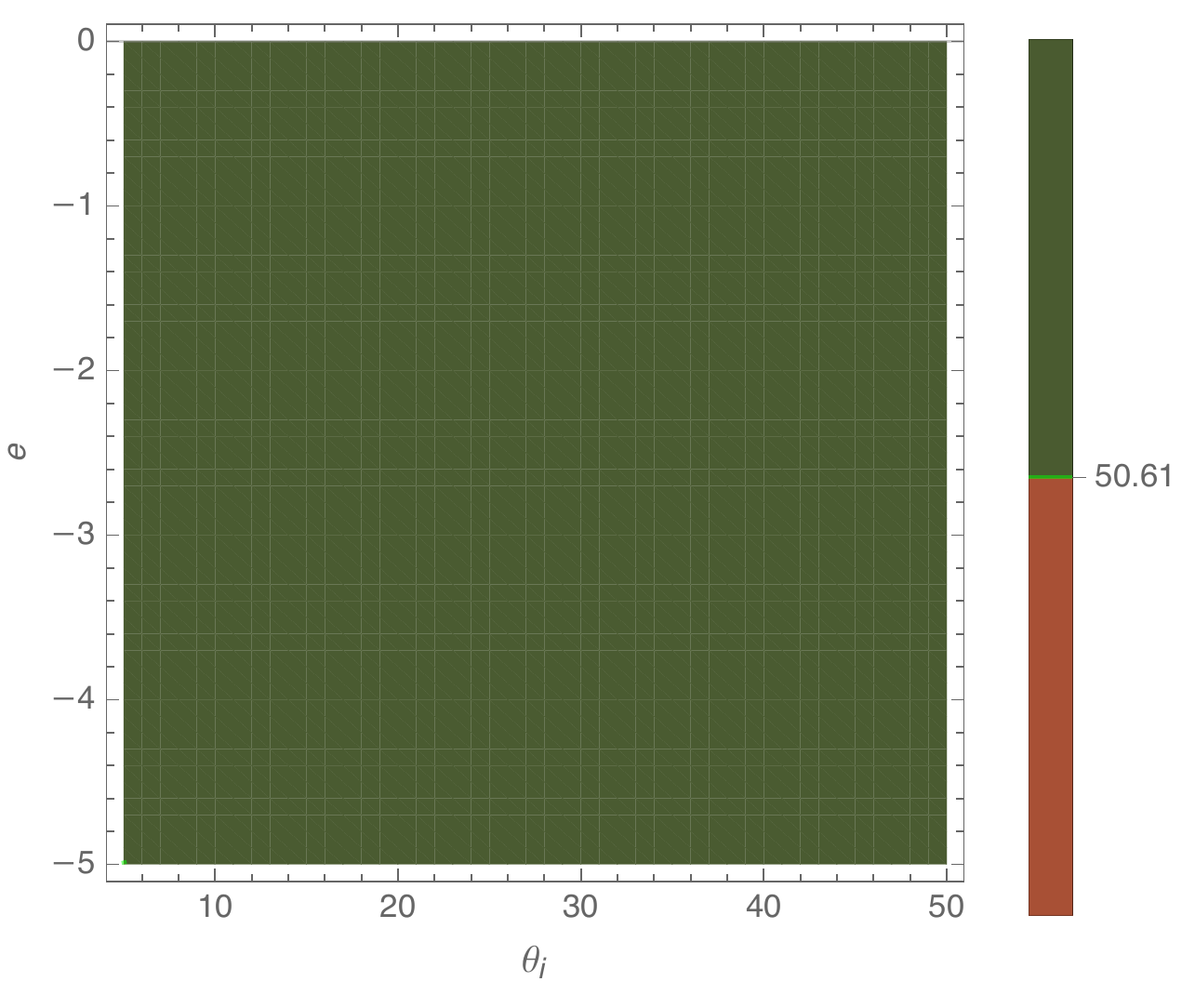}
	   
	\end{minipage}}
 \hfill 	
  \subfloat[For $0 \leq e \leq 1.68$]{
	\begin{minipage}[c][1\width]{
	   0.3\textwidth}
	   \centering
	   \includegraphics[width=1.1\textwidth]{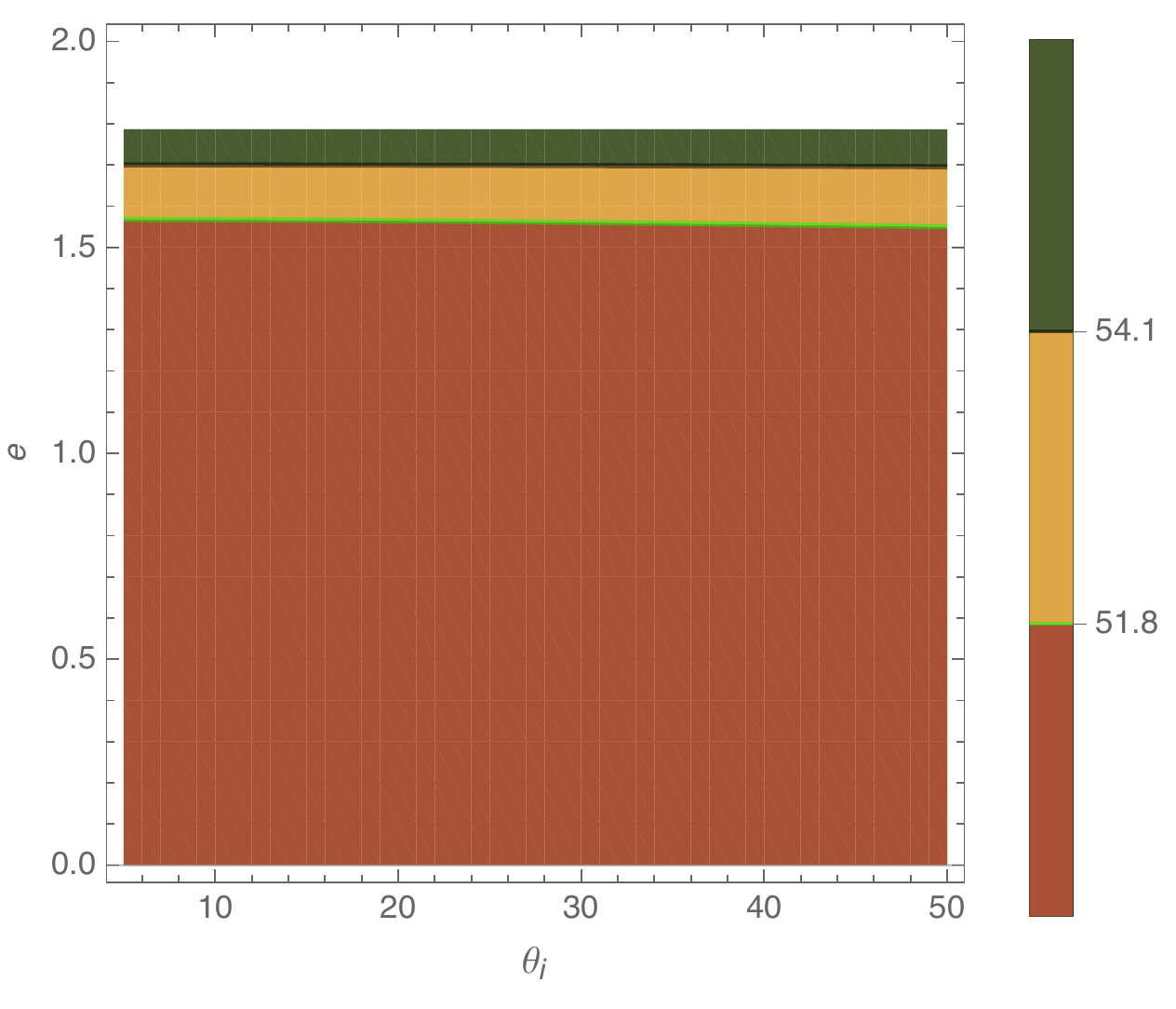}
	\end{minipage}}
 \hfill	
  \subfloat[For $1.68 \leq e \leq 2$]{
	\begin{minipage}[c][1\width]{
	   0.3\textwidth}
	   \centering
	   \includegraphics[width=1.1\textwidth]{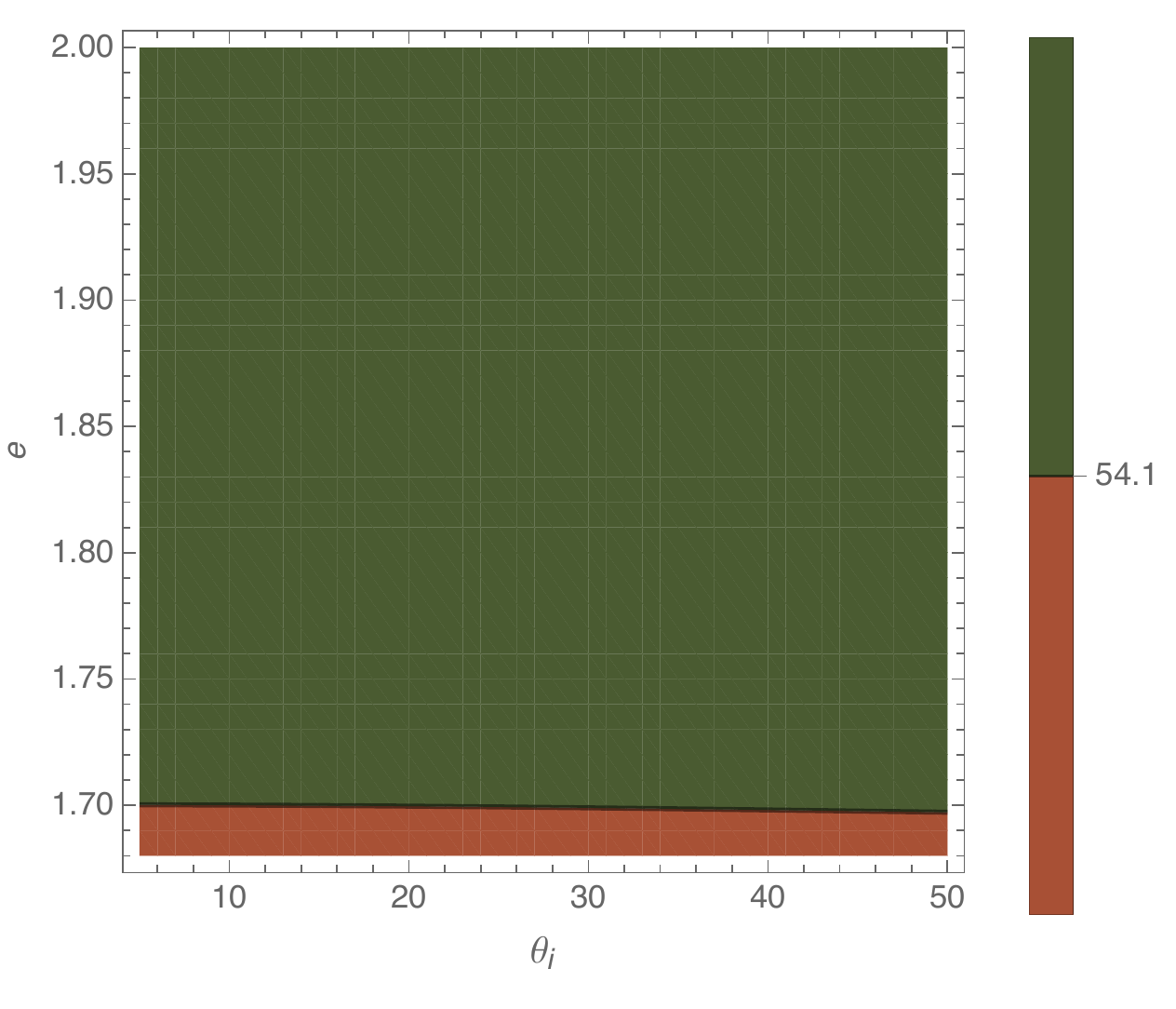}
	\end{minipage}}
\caption{Contour plots for angular size of $Sgr\, A^*$ shadow if modelled by the metric in Eq.(\ref{g}). Values of $\theta_i \in [5,50]$, and $e \in (-5,2)$ for which the shadow size is less than $49.5\, \mu$as or greater than $54.1\, \mu$as are disallowed. Value of $b$ does not effect the shadow size.}
\label{SgrAplt}
\end{figure}

\noindent
For the case of $Sgr\, A^*$, a unique spin parameter $a_* \approx 0.44$ can be fixed by studying the multiple quasi-periodic oscillations of radio emissions \cite{Miyoshi, Kato_2010}. However, EHT shadow observation is consistent with all inclination angles that are less than $50 ^{\circ}$. Therefore, we can show the variation in shadow size using contour plots for different choices of $(\theta_i, e)$, see Fig.[\ref{SgrAplt}]. It is interesting to note that the $Sgr\, A^*$ shadow size is consistent with EHT observations for all values of $e \in \Big(-3 - 3/ \sqrt{1-(0.44)^2} \approx -6.341, 1.7\Big]$ with $\theta_i < 50^{\circ}$ and $a_* \approx 0.44$. Whereas all values of $e > 1.7$ are disallowed by EHT observations \cite{EventHorizonTelescope:2022xnr}.\\

\section{Discussion and Conclusion}
BHs are inevitable consequence of Einstein's field equations. The recent observation of shadow by the EHT along with the detection of gravitational wave by the LIGO-Virgo-KAGRA (LVK) collaboration provide direct and compelling evidence for the existence of such objects in nature. These astrophysical observations are in consistent with the Kerr paradigm, namely the central compact object being a Kerr BH completely described by the mass (M) and spin (a). However, as discussed in the introduction, the Kerr solution suffers from various pathologies such as the existence of spacetime singularities and mass inflation singularity at the Cauchy horizon. It is widely believed that such issues are consequence of the limitations of classical GR and will be resolved by a full quantum gravity formalism. However, in the absence of a consistent quantum theory of gravity, one may try to tackle the these problems using phenomenological models. One such model of regularized stable Kerr BH has been recently developed in Refs. \cite{Carballo-Rubio:2022kad, Franzin:2022wai}. In addition to mass and spin, these BH solutions depend on two additional parameters: a conformal factor $b$, and a deviation parameter $e$ that regularizes and stabilizes the Kerr solution. In this work, we have studied some observational signatures of these parameters on BH QNMs and shadow observations. \\
 
\noindent
Astrophysical BHs are expected to be stable under perturbations. As a result, stability turns out to be an important criteria for any well behaved BH spacetime. One of the main goal of this work is to put bounds on $(b,e)$ requiring stability of regularized stable Kerr BH when perturbed. For this purpose, we compute the scalar QNM modes and investigate whether the imaginary part of  these modes are negative needed for the stability. Our analysis shows that the regularized Kerr BHs are stable under scalar perturbation for all allowed values of $(b,e)$. Interestingly, for a fixed value of $e$, the imaginary parts $\omega_I$ of QNM frequencies decrease as we increase the values of $b$, which makes the BHs ``more" stable as it will take less relaxation time to shake off the perturbation. Also, for a fixed $a < M$, in the ``e-extremal" limit $(e \to 2)$, both imaginary and real parts of QNM frequencies tend towards zero. This gives us a way to identify ``e-extremality". Since small values $\omega_I$ lead to longer relaxation time, near-extremal values of $e$ make the BH ``less" stable.\\

\noindent
We have also studied the shadow structure of these regularized BH and its dependence on the additional parameters $(e,b)$. Since the parameter $b$ appears in the conformal factor, it doesn't affect the null geodesics and the shadow radius. In contrast, choice of $e$ values has a sharp impact on the angular shadow size, which we compare with $M87^*$ and $Sgr\, A^*$ images. Consistency with EHT shadow observations put stringent restrictions on the parameter space of $e$, but keep $b$ unconstrained. For the allowed values of the $e$, these regularized BHs mimic the Kerr shadow of corresponding mass and spin, but can still be distinguished from Kerr using QNM structure. To the best of our knowledge, this is the first observational bounds on the parameters of regularized BHs in consideration.\\

\noindent
In future, one may try to extend the consistency criteria discussed in Sec.\ref{Sec_3} on the mass profile $m(r)$ by taking back reaction into account. Also, we want to study other important properties of these BHs, such as their response to an external tidal field and check whether these BHs have zero Love number like Kerr. Further constraints in the parameter space can come from the study of gravitational lensing as well.\\

\section*{Acknowledgement}
The research of R.G. is supported by the Prime Minister Research Fellowship (PMRF Reference Number: 192002-120 and ID: 1700531), Government of India. Research of M.R. is supported by SERB, Government of India through the National Post Doctoral Fellowship grant (Reg. No. PDF/2021/001234). Research of AKM is supported by SERB, Government of India through the National Post Doctoral Fellowship grant (Reg. No. PDF/2021/003081). R.G. wants to thank Sudipta Sarkar for many useful discussions.

\bibliography{Draft_v1}

\bibliographystyle{./utphys1}
\end{document}